\documentclass[
    ,final            
  ]
  {aipproc}

\layoutstyle{6x9}
\usepackage{amsmath}
\usepackage{lscape,colordvi}
\def\vspa{\vspace{-1cm}}
\def\vspaa{\vspace{-.7cm}}

\def\vspbb{\vspace{-.35cm}}
\def\vspb{\vspace{-.5cm}}
\def\vspc{\vspace{-.4cm}}

\def\etp{\enlargethispage{.64cm}}
\usepackage{verbatim}

\begin{document}

\title[Le Delliou and Mimoso: Separating expansion from contraction \dots]{Separating expansion from contraction and generalizing TOV condition
in spherically symmetric models with pressure}


\author{Morgan Le Delliou}{
  address={\textbf{Speaker};Instituto de F\'isica Te\'orica UAM/CSIC, Facultad
de Ciencias, C-XI, Universidad Aut\'onoma de Madrid, Cantoblanco, 28049
Madrid SPAIN. Email: Morgan.LeDelliou@uam.es},
,altaddress={Centro de F\'isica Te\'orica e Computacional, Universidade de Lisboa,  Av. Gama Pinto 2, 1649-003 Lisboa, Portugal.} 
}

\author{Jos\'e P. Mimoso}{
  address={Departamento de F\'isica, Faculdade de Ci\^encias, Edif\'icio C8, Campo
Grande, P-1749-016 Lisboa, Portugal Email: jpmimoso@cii.fc.ul.pt},
,altaddress={Centro de F\'isica Te\'orica e Computacional, Universidade de Lisboa,  Av. Gama Pinto 2, 1649-003 Lisboa, Portugal.}
}


\begin{abstract}
We investigate spherically symmetric solutions with pressure and discuss
the existence of a dividing shell separating expanding and collapsing
regions. We perform a 3+1 splitting and obtain gauge invariant conditions
relating not only the intrinsic spatial curvature of the shells to
the ADM mass, but also a function of the pressure which we introduce
that generalises the Tolman-Oppenheimer-Volkoff equilibrium condition.
We consider the particular case of a Lema\^itre-Tolman dust models
with a cosmological constant (a $\Lambda$-CDM model) as an example
of our results. 
\end{abstract}

\maketitle

\section{Introduction\vspc}

Cosmological formation of structure assumes the collapse of inhomogeneities,
via gravitational instability, into \char`\"{}bound\char`\"{} structures,
with the underlying idea that they depart from the cosmological expansion.
This approach usually models overdensities with closed patches embedded
in Friedman backgrounds, in particular in the spherical collapse included
in the Press \& Schechter scheme \cite{PS74}. Birkhoff's theorem
is invoked to justify their independent evolutions \cite{PeeblesPadm},
while it is proved for asymptotically flat spacetimes \cite{Birkhoff23}.
Such separation is reminiscent of the concept of trapped surfaces
\cite{TrappedSurf}.

In this work we define general conditions for the existence of a shell
separating contraction from expansion, and will illustrate our results
with a simple example of perturbed $\Lambda$-CDM models. We adopt
the Generalised Painlev\'e-Gullstrand (hereafter GPG) formalism used
in Lasky \& Lun \cite{LaskyLun06b}, which involves a $3+1$ splitting
(ADM) and the consideration of gauge invariants kinematic quantities
\cite{EllisElst98}.
\vspb

\section{ADM approach to LTB models in GPG system\vspc}

\etp
We consider a spherically symmetric Generalised Lema\`itre-Tolman-Bondi
(Generalised LTB, or GLTB) metric to include pressure. Performing
an ADM 3+1 splitting in the GPG coordinates~\cite{LaskyLun06b} ,
the metric reads\vspace{-0.38cm}
\begin{align}
ds^{2} & =-\alpha(t,r)^{2}dt^{2}+\frac{1}{1+E(t,r)}\left(\beta(t,r)dt+dr\right)^{2}+r^{2}d\Omega^{2}\;,\end{align}
\vspace{-0.5cm}

\noindent characterised where $\alpha$ is a lapse function, $\beta$
is a shift, and $E$ is a curvature-energy characterising the curvature
of the spatial surfaces orthogonal to the direction $n_{a}=(-\alpha,0,0,0)$
of the flow. For a perfect fluid, the projected Bianchi identities
$T_{b;a}^{a}=0$ yield the energy density conservation equation when
we project along the flow $n^{b}$, and the Euler equation when we
project orthogonally to the latter. Using the projection $h_{a}^{\, b}$,
and denoting the density Lie derivative along the flow $\mathcal{L}_{n}\rho$,
we have:\vspace{-0.55cm}
\begin{align}
n^{b}T_{b;a}^{a}= & -\mathcal{L}_{n}\rho-\left(\rho+P\right)\Theta=0, & h_{a}^{\, b}T_{b;c}^{c}= & 0\Rightarrow P^{\prime}=-\left(\rho+P\right)\frac{\alpha^{\prime}}{\alpha}\;,\label{eq:Bianchi}\end{align}

\vspace{-0.3cm}

\noindent where $P$ is the pressure, the radial derivatives are denoted
by a prime, $^{\prime}$, the time derivatives are represented by
a dot, $\dot{}$, and the expansion is $\Theta$.

Introducing the ADM (also called Misner-Sharp) mass\vspace{-0.55cm}
\begin{align}
M & =r^{2}\left(1+E\right)\left(\ln\alpha\right)^{\prime}-4\pi Pr^{3}+\frac{1}{3}\Lambda r^{3}+r^{2}\mathcal{L}_{n}\left(\frac{\beta}{\alpha}\right)\;,\end{align}
\vspace{-0.5cm}

\noindent where $\Lambda$ is the cosmological constant, we write
the Einstein's field equations (EFEs) as Lie derivatives along the
flow 
\vspace{-0.26cm}
\begin{align}
\mathcal{L}_{n}{E}= & 2\left(\frac{\beta}{\alpha}\right)\frac{1+E}{\rho+P}P^{\prime}, & \Rightarrow\dot{E}= & \beta\left(E^{\prime}+2\frac{1+E}{\rho+P}P^{\prime}\right),\\
\mathcal{L}_{n}M= & 4\pi Pr^{2}\left(\frac{\beta}{\alpha}\right), & \Rightarrow\dot{M}= & \beta\left(M^{\prime}+4\pi Pr^{2}\right),\\
\left(\frac{\beta}{\alpha}\right)^{2}= & E+2\frac{M}{r}+\frac{1}{3}\Lambda r^{2}.\label{eq:radial}\end{align}
\vspace{-0.48cm}

\noindent %
\begin{figure}
\begin{centering}
\includegraphics[bb=146bp 342bp 821bp 683bp,clip,width=0.4\paperwidth]{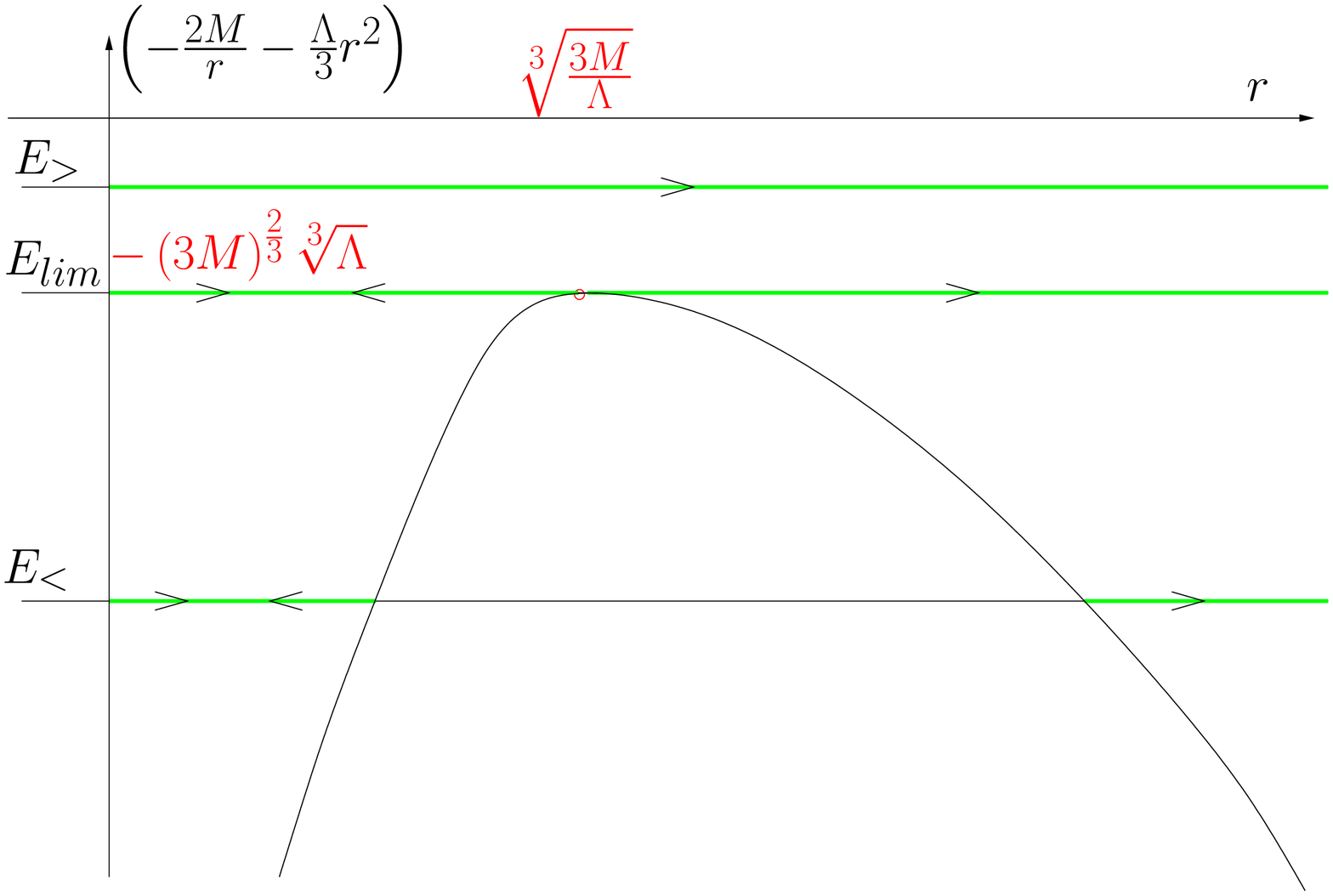}
\par\end{centering}

\caption{\label{fig:KinAn}kinematic analysis of motion in the pseudo-potential
V}

\end{figure}
The system becomes then closed when an equation of state (EoS) $f(\rho,P)=0$
is supplied. The $\Lambda$ term can be\textbf{ }absorbed and many
fluids mass equations written with each component using the $\frac{\beta}{\alpha}$
term for the overall sum of the masses.\vspb

\section{Separating collapse from expansion\vspc}

\noindent \etp To characterise the separation of expansion from collapse,
we prefer to use here GLTB coordinates in which the flow direction
reduces to time by choosing $\beta=-\dot{r}$. We thus have\vspace{-0.35cm}
\begin{align}
ds^{2} & =-\alpha(T,R)^{2}\left(\partial_{T}t\right)^{2}dT^{2}+\frac{\left(\partial_{R}r\right)^{2}}{1+E(T,R)}dR^{2}+r^{2}d\Omega^{2},\\
\dot{M}= & \beta4\pi Pr^{2},\,\dot{E}r^{\prime}=2\beta\frac{1+E}{\rho+P}P^{\prime},\,\left(-\frac{\dot{r}}{\alpha}\right)^{2}=E+2\frac{M}{r}+\frac{1}{3}\Lambda r^{2}.\label{eq:radialLTB}\end{align}

\vspace{-0.34cm}

There are two situations one should consider in parallel. On the one
hand, we look for the vanishing of the gauge invariant expansion $\Theta$,
defined as ${n^{a}}_{;a}$, since our goal is to separate an inner
collapsing, spherical region from the outer expanding universe. On
the other hand, the total ADM mass of the spherical region that departs
from the expansion flow should be conserved. This is indeed suggested
by the dust case where that happens for every shell.

We denote with $\star$ the dividing shell. The analysis of the kinematic
quantities reveals that $\Theta$ is linked to the shear $a$, in
that at the $\star$ shell\vspace{-0.3cm}
\begin{align}
\Theta= & -3\left(a+\frac{\beta}{\alpha}\frac{1}{r}\right)\Leftrightarrow & \Theta_{\star}+3a_{\star}= & -3\left.\frac{\beta}{\alpha}\right|_{\star}\frac{1}{r_{\star}}\left(=0\textrm{ when }\mathcal{L}_{n}M(t,r_{\star}(t))=0\right).\end{align}
\vspace{-0.55cm}

\noindent Moreover the generalized Friedman constraint\vspace{-0.3cm}
\begin{align}
^{(3)}R+\frac{2}{3}\Theta^{2}= & 6a^{2}+16\pi\rho+2\Lambda\end{align}
\vspace{-0.55cm}

\noindent tells us that the vanishing of $\Theta$ only happens in
region of positive 3-curvature $^{(3)}R$. On the other hand, if we
demand that the separating shell has a dust-like vanishing mass/energy
flow, i.e., has a conserved ADM mass along $n^{a}$:\vspace{-0.5cm}
\begin{align}
\forall t,\,\mathcal{L}_{n}M(t,r_{\star}(t))= & 0 & \Leftrightarrow\forall t,\, E= & -2\frac{M}{r_{\star}}<0.\label{eq:Estar}\end{align}
\vspace{-0.6cm}

\noindent We remark that $M$ refers to the total ADM mass, thus including
a cosmological constant. 

Taking into account that the equilibrium of static spherical configurations
requires the satisfaction of the Tolman-Oppenheimer-Volkoff equation
of state \cite{TOV}, we are led to define a generalized TOV function\vspace{-0.66cm}

\begin{align}
\mathrm{TOV}= & \left[\frac{1+E}{\rho+P}P^{\prime}+4\pi Pr+\frac{M}{r^{2}}-\frac{1}{3}\Lambda r\right]=\mathcal{L}_{n}\left(\frac{\beta}{\alpha}\right)\;,\label{eq:TOVdef}\end{align}

\vspace{-0.27cm}
which reduces to the usual TOV equation when it vanishes.

The radial behaviour of the $\star$-shell is then similar to a turnaround
shell. Indeed $r_{\star}=-\frac{2M_{\star}}{E_{\star}}$ leads to
a null radial velocity $\dot{r}_{\star}=0$ while its acceleration
reveals the importance of the TOV parameter\vspace{-0.35cm}
\begin{align}
\ddot{r}_{\star}= & -\alpha^{2}\left[\mathrm{TOV}_{\star}-r_{\star}^{2}\frac{\mathrm{TOV}_{\star}^{2}}{M_{\star}}\right]; & \ddot{r}_{LTB,\star}= & -\alpha^{2}\mathrm{TOV}_{\star}.\end{align}
\vspace{-0.5cm}

\noindent The local staticity of the $\star$-shell is then shown
to be equivalent to having a local TOV equation on this limit shell\vspace{-0.5cm}
\begin{align}
\mathrm{TOV}_{\star}= & 0 & \Leftrightarrow-\frac{1}{\rho+P}P^{\prime}= & \left[\frac{4\pi Pr+\frac{M}{r^{2}}}{1-\frac{2M}{r}}\right]_{\star}.\end{align}
\vspa

\section{An example: $\Lambda$CDM\vspc}

A simple illustration of our result is given by the case of dust with
a $\Lambda$. There is then no pressure gradients and $M_{dust}$
and $E$ are conserved, which simplify the analysis (i.e. $\alpha=1$),
and allow us to perform a kinematic analysis (see Fig. \ref{fig:KinAn})
per shell of Eq. (\ref{eq:radialLTB}-c). An effective potential is
defined by $V(r)\equiv-\frac{2M}{r}-\frac{\Lambda}{3}r^{2}$, so that
when $\dot{r}=0$ we have indeed $V(r)=E$:\etp\vspace{-0.4cm}
\begin{align}
 &  & \dot{r}^{2}= & 2\frac{M}{r}+\frac{1}{3}\Lambda r^{2}+E, & \textrm{with }\ddot{r}= & -\frac{M}{r^{2}}+\frac{\Lambda}{3}r.\end{align}
\vspace{-0.55cm}

\noindent For each shell there is a virtual static state at $r_{lim}=\sqrt[3]{\frac{3M}{\Lambda}}$,
$E_{lim}=-\left(3M\right)^{\frac{2}{3}}\Lambda^{\frac{1}{3}}$, which
only depends on $\Lambda$ and $M(R)$. We also recall that, in this
model, $\mathrm{TOV}=\frac{M}{r^{2}}-\frac{\Lambda}{3}r=-\ddot{r}$. 

We then need to choose initial conditions $\rho_{i}$, which sets
the $E_{lim}$ profile, and $v_{i}$, which sets the $E_{i}$ profile.
The intersection of $E_{lim}$ with $E_{i}$ will be static at turnaround,
hence defining our limit shell. We use two sets of cosmologically
motivated initial conditions: one with $\rho_{i}$ as an NFW profile
\cite{NFW} and $E_{i}$ as a parabola, and the second,  with a cuspless
power law as $\rho_{i}$ and a Hubble flow for $v_{i}$.

\vspaa

\section{Conclusions\vspc}


Using non-singular, Generalized Painlev\'e-Gullstrand coordinate
formulation of the ADM spherically symmetric, perfect fluid system
\cite{LaskyLun06b} we have shown \cite{MimLeD} that the existence
of shells locally separating between inner collapsing and outer expanding
regions, is governed by the condition that the combination of expansion
scalar and shear $\theta+3a$ should vanish on the shell. The ADM
mass of the shell is then conserved. This condition requires that
the separating shell must be located in an elliptic ($E<0$) region.
Moreover, for that shell to exist for certain over time, we have shown
that the TOV equation must be locally satisfied. 
We argue that this local condition is global in a cosmological context
(FLRW match at radial asymptote). Given appropriate initial conditions,
this translates into global separations between an expanding outer
region and an eventually collapsing inner region. We present simple
but physically interesting illustrations of the results, a model of
Lema\^itre-Tolman dust with $\Lambda$ representing spherical perturbations
in a $\Lambda$CDM with two different initial sets of cosmologically
interesting conditions consistent with known phenomenological constraints
\cite[and Refs. therein]{MimLeD}: an NFW density profile with a simple
curvature profile going from bound to unbound conditions, and a non
cuspy power law fluctuation with initial Hubble flow. We show, for
these models, the existence of a global separation. We argue that
these shells are trapped matter surfaces \cite{MimLeD} and that they
constitute the validity locus to an analog to Birkhoff's theorem.
\vspbb\etp


\begin{theacknowledgments}

\vspc

The work of MLeD is supported by CSIC (Spain) under the contract JAEDoc072,
with partial support from CICYT project FPA2006-05807, at the IFT,
Universidad Autonoma de Madrid, Spain, and was also supported by FCT
(Portugal) under the grant SFRH/BD/16630/2004, at the CFTC, Lisbon
University, Portugal. Financial support from the Foundation of the
University of Lisbon and the Portuguese Foundation for Science and
Technology (FCT) under contracts POCI/FP/ FNU/50216/2003 and POCTI/ISFL/2/618
is also gratefully acknowledged. The authors are grateful to Filipe
Mena for many helpful discussions\vspb

\end{theacknowledgments}





\IfFileExists{\jobname.bbl}{}
 {\typeout{}
  \typeout{******************************************}
  \typeout{** Please run "bibtex \jobname" to optain}
  \typeout{** the bibliography and then re-run LaTeX}
  \typeout{** twice to fix the references!}
  \typeout{******************************************}
  \typeout{}
 }

\end{document}